\documentclass[aps,prd,twocolumn,superscriptaddress,nofootinbib,showpacs,preprintnumbers]{revtex4}
\usepackage{graphicx}
\usepackage[latin1]{inputenc}

\begin{document}
\preprint{IFF-RCA-09-10-08}
\title{Is the 2008 NASA/ESA double Einstein ring actually a
ringhole signature?}

\author{Pedro F. Gonz\'{a}lez-D\'{\i}az}
\email{p.gonzalezdiaz@imaff.cfmac.csic.es } \affiliation{Colina de
los Chopos, Instituto de F\'{\i}sica Fundamental, \\
Consejo Superior de Investigaciones Cient\'{\i}ficas, Serrano 121,
28006 Madrid, Spain}

\begin{abstract}
It is argued that whereas the Shatskiy single rings produced by
the gravitational inner field of a spherically symmetric wormhole
could not be used to identify its presence in the universe or the
contents of a parallel universe because such rings may be confused
with the most familiar Einstein rings, the image which the inner
gravitational field of a ringhole with toroidal symmetry would
allow us to detect from a single luminous source situated behind
the ringhole in our universe or in a parallel universe is that of
two concentric bright rings, and this is a signature that cannot
be attributed to any other single astronomical object in whichever
universe it may be placed. At the beginning of 2008 the NASA/ESA
Hubble Space Telescope revealed a never-before-seen phenomenon in
space: a pair of glowing rings, one nestled inside the other like
a bull's-eye pattern. It is our alternate proposal in this paper
to attribute such a discovery to the first astronomical ringhole
found in the universe, rather than to the highly unlikely double
lensing effect produced by the required ultra precise alignment of
three galaxies along the line of sight. After all, a ringhole 
is a perfectly valid solution to the Einstein equations and the 
stuff which makes it possible is becoming more and more familiar 
in cosmology.
\end{abstract}


\maketitle

Among the reactions that greeted the Shatskiy proposal [1] that
wormholes, which are usually disguised as black holes, can be made
observable and recognizable in terms of bright, glowing rings
originating from the necessary flaring out which is produced by
the presence of the so called phantom matter around their throat,
there was one typical outburst which was remarked by the sentence:
"It is an interesting thing to think about, maybe after a few
beers." [2]. For sure, even more offensive statements were
expressed against the existence of black holes some 30 years ago.
As we all now know black holes have become commonplace in
astronomy and fundamental physics. Similarly, wormholes are
expected by an increasing number of scientists to also become
commonplace in physics not too far in the future. There were more
serious criticisms to the Shatskiy work though. In my opinion, the
really most devastating argument against the wormhole
distinguishable character of the Shatskiy rings is that, even if
exotic matter does exist, other many objects are able to create a
similar light signature [3]. In particular, it is hard to see how
these rings could be differentiated from the astronomical
blueprint left by negative energy stars and, mainly, from all
those massive astronomical objects whose gravitational lensing
effects appear as the so called Einstein rings.

The actual problem is with the symmetry of the throat. Wormholes
are characterized by a spherically symmetric throat and,
therefore, the diverging lensing effect would necessarily manifest
by the observer interpretation of the luminous source as a single
ring source, as indicated by Shatskiy. This pattern could well be
misinterpreted as being originated from a star or other massive
astronomical object, instead of a wormhole, with a radius quite
smaller than that for that wormhole throat radius. An inner
tunneling symmetry which would give rise to an inexorably
distinguishable lensing pattern is that of a ringhole [4], that
is, a space-time tunnel whose throat has the toroidal symmetry
(see Fig. 1 (a)). Using the set of geometrical parameters
specified in this upper part of Fig. (1) we can derive the metric
for a ringhole to be [4]
\begin{equation}
ds^2= -C_2 r^2 dt^2 +b^2\left[1+\frac{C_1 a^2\sin^2\varphi_2}{r^6\left(1-\frac{A^2}{r^4}\right)}\right]d\varphi_2^2 + m^2 d\varphi_1^2
\end{equation}
where
\begin{equation}
A=a^2-b^2,\;\; m=a-b\cos\varphi_2 ,\;\; r=\sqrt{a^2+b^2-2ab\cos\varphi_2},
\end{equation}
with $C_1$ and $C_2$ arbitrary integration constants, and $a$ and $b$ the radius
of the circumference generated by the circular axis of the torus and that of a
torus section, respectively, with $a>b$. Metric (1) is defined for $0\leq t\leq\infty$,
$a-b\leq r\leq a+b$ and the angles (see Fig. 1 (a)) $0\leq \varphi_1$, $\varphi_2 \leq 2\pi$.

In order to check the properties of a ringhole as a lens, we now
write the static spacetime metric of a single, traversible
ringhole in the form
\begin{equation}
ds^2 =-dt^2 +\left(\frac{n_{\ell}}{r_{\ell}}\right)^2 d\ell^2
+m_{\ell}^2 d\varphi_1^2+\left(\ell^2+b_0^2\right)d\varphi_2^2 ,
\end{equation}
where $-\infty <t < +\infty$, with $-\infty <\ell < +\infty$ the
proper radial distance of each transversal section of the torus,
and
\begin{equation}
m_{\ell}=a-\left(\ell^2 +b_0^2\right)^{1/2}\cos\varphi_2,\;\;
n_{\ell}=\left(\ell^2 +b_0^2\right)^{1/2}-a\cos\varphi_2 ,
\end{equation}
\begin{equation}
r_{\ell}=\sqrt{a^2 +ell^2 +b_0^2 -2\left(\ell^2
+b_0^2\right)^{1/2}a\cos\varphi_2} ,
\end{equation}
in which $b_0$ is the throat radius. As $\ell$ increases from
$-\infty$ to 0, $b$ decreases monotonously from $+\infty$ to its
minimum value $b_0$ at the throat radius, and as $\ell$ increases
onward to $+\infty$, $b$ increases monotonously to $+\infty$
again. Now, for metric (3) to describe a ringhole we must embed it
in a three-dimensional Euclidean space at fixed time $t$ [4] whose
metric can be written as
\begin{equation}
ds^2=dz^2 +dr^2 +r^2 d\phi^2
=\left[1+\left(\frac{dz}{dr}\right)^2\right]dr^2 +r^2 d\phi^2 ,
\end{equation}
with $dz/dr=\left(b^2/b_0^2 -1\right)^{-1/2}$. The requirement
that ringholes be connectible to asymptotically flat spacetime
entails at the throat that the embedding surface flares outward
for $2\pi - \varphi_2^c >\varphi_2 >\varphi_2^c$, and flares
inward for $- \varphi_2^c <\varphi_2 <\varphi_2^c$, with
$\varphi_2^c =\arccos(b/a)$, which respectively satisfy the
condition $d^2 r/dz^2 >0$ and $d^2 r/dz^2 < 0$ at or near the
throat.

It follows [4] that one would expect lensing effects to occur at
or near the ringhole throat, that is to say, the mouths would act
like a diverging lens for world lines along $2\pi - \varphi_2^c
>\varphi_2 >\varphi_2^c$, and like a converging lens for world
lines along $- \varphi_2^c <\varphi_2 <\varphi_2^c$. No lensing
actions would therefore take place at $\varphi_2=\varphi_2^c$ and
$\varphi_2=2\pi - \varphi_2^c$.

\begin{figure}
\includegraphics[width=.9\columnwidth]{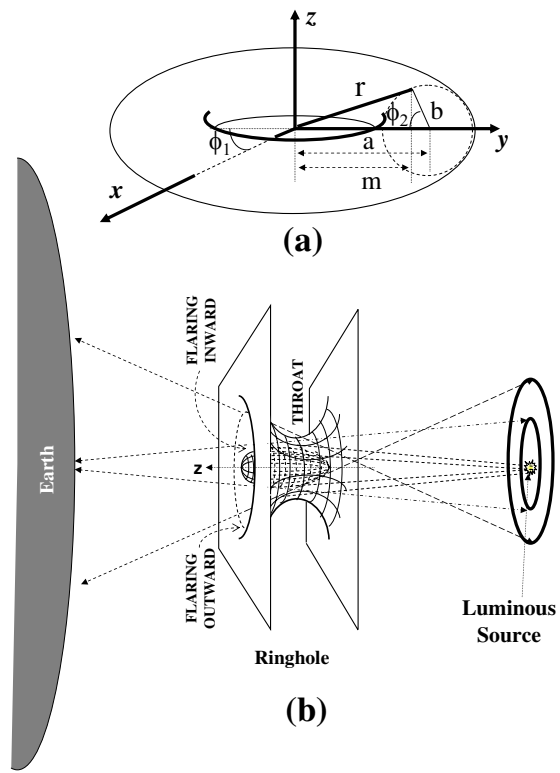}
\caption{\label{fig:epsart} Gravitational lensing effect produced
by a ringhole from a single luminous source. (a) Parameters
defining the toroidal ringhole throat in terms of which metric (1)
is defined. (b) Rays passing near the outer and inner surfaces
respectively flare outward and inward, leading to a image from a
luminous point placed behind the ringhole which is made of two
concentric bright rings. The relative mutual positions of these
rings would depend on the distance between the ringhole and the
luminous source. If that distance is small enough then the larger
outer ring comes from the flaring inward surface, and conversely,
if the distance source-ringhole is increased then the outer ring
comes from the outward surface, the larger that distance the
greater the difference between the two bright rings.}
\end{figure}

In fact, in the case of ringholes, instead of producing just a
single flaring outward for light rays passing through the wormhole
throat, this multiply connected topology, in addition to that
flaring outward (diverging) effect, also produces a flaring inward
(converging) effect [4] on the light rays that pass through its
throat, in such a way that an observer on Earth would interpret
light passing through the ringhole throat from a single luminous source
as coming from two bright, glowing concentric rings, which forms up the
distinctive peculiar pattern from ringholes (See Fig. 1 (b)). That
pattern cannot be generated by any other possible disturbing
astronomical object other than a very implausible set of three
luminous massive objects (let us say galaxies) which must be so
perfectly aligned along the sigh line that its occurrence becomes
extremely unlikely.

It is readily inferred from Fig. 1 (b) that, for a reasonably
large ringhole sufficiently far from the luminous source, the
inner bright ring would correspond to the flaring inward
(converging) surfaces. If we keep the ringhole size invariant and
the distance between the ringhole and the luminous source is
decreased drastically, then the inner bright ring would turn to be
produced by the flaring outward (diverging) surface.

Such a ringhole signature may have been already observed though it has 
been so far attributed to the combined effect of two Einstein rings
originated from the above-considered to be extremely unlikely
superprecise alignment of three galaxies. In fact, at the
beginning of 2008 The NASA/ESA Hubble Space Telescope revealed [5]
a never-before-seen phenomenon in space: a pair of glowing rings,
one nestled inside the other like a bull's-eye pattern. This
double-ring pattern was interpreted as a double Einstein ring
being caused by the complex bending of light from two distant
galaxies strung directly behind a foreground massive galaxy, like
three beads on a string along the line of sight, 
simply because at the time there were no
other available interpretations for what was being observed. Being
more than just a novelty, this very rare phenomenon found with the
Hubble Space Telescope could, moreover, eventually offer insight
into dark matter, dark energy, the nature of distant galaxies, and
even the curvature of the Universe.

As previously stated, for that interpretation to be feasible, 
the massive foreground galaxy had to be
almost perfectly aligned in the sky with two background galaxies
at different distances to justify the finding. The foreground
galaxy is 3 billion light-years away. Now, in order to justify the
ratio between the two ring radii, the inner ring and outer ring
would be comprised of multiple images of two galaxies  at a
distance of some 6 billion and approximately 11 billion
light-years.

However, the odds of observing the required extremely precise
alignment of the three galaxies are so small (an estimated 1 in
10,000) that even some of the discoverers of that astronomical
phenomena said that they had 'hit the jackpot' with the discovery.
At the time, the authors of Ref. [5] had no alternative other than
accepting that quite improbable interpretation of the result.
Nevertheless, having we uncovered in this letter that such
concentric rings may well be also interpreted as the blueprint of
the presence of a ringhole in the direction in space where the
double bright ring system was discovered, we adopt the latter
interpretation in terms of a ringhole as the most probable
explanation for that phenomenon, taking now the luminous sources at 
redshifts corresponding to 3 and 6 billion light-years as measuring 
the positions of the two ringhole mouths on the sky, and their 
respective luminosities as stemming from the respective light 
deflections along the angle $\varphi_2$ caused by the combined effect
of the size of the throat radius and the relative distance between
the two mouths.

In this case, besides valuable information on dark matter, dark energy 
and universe curvature, what could eventually be most astonishing in 
its implications would be an unprecedented insight into the content of other
universes linked to ours by means of ringholes. Not with standing,
in spite of the apparent evidence in its favor, I
only present here the ringhole interpretation of the results of
Ref. [5] as just a possible alternate implication, probably the
most likely one now at our disposal. After all, a ringhole is a
perfectly valid solution to the Einstein equations for a exotic
stuff - possibly phantom energy- which is becoming more and more
familiar in the full context of current cosmology. The potentially
attainable insight from such an interpretation is twofold. On the
one hand, we would get a direct evidence for the existence of
ringholes and, by the way, possibly of wormholes, and on the other
hand, we could have found the door to a parallel universe, and
hence got a first direct evidence for the existence of the
multiverse scenario.

There is an observation which may in principle distinguish a
static ringhole staying within our own universe and having its two
mouths at rest with respect to each another, from a ringhole that
connect our universe to a parallel universe or, in general, to
other universe of a multiverse scenario. In the latter case since
there is no common space-time for the two universes (parallel or
not), the two mouths should necessarily be in perpetual quasi
periodic relative random motion with completely unspecified speed.
This would make the time and space for the two universes at all
independent because the relative motion of the two mouths converts
the ringhole in a time machine that contains completely arbitrary
closed timelike curves. In the case of the inner static ringhole,
if the luminous source is kept motionless and the ringhole does
not behave like a time machine, the two concentric rings would be
well resolved and defined on the pattern. However, if the
positions of the two mouths continuously vary relative to each
other in a random though quasi periodic way then the width of each of the
two concentric rings would be stretched out and their resolution
spoiled and clearly blurred due to the continuous and completely
arbitrary changes of distance between the two mouths, thus leading
to a glowing background around the rings, showing just a maximum
of intensity at the average relative position of the mouths,
provided the relative motion keep a sufficiently high degree of
periodicity. In the latter case, the metric of the ringhole would
change to be given by a line element that describes arbitrary time
travel induced by a nearly periodic relative motion between the
two mouths. Using arguments similar to those used in Ref. [4] 
we finally get
\begin{equation}
ds^2=-\left[1+\bar{g}F(\ell)\sin\varphi_1\right]^2 dt^2
+d\ell^2+m_{\ell}^2 d\varphi_1^2 +b^2 d\varphi_2^2 ,
\end{equation}
where $\bar{g}=\bar{\gamma}^2\frac{d\bar{v}}{dt}$ is the average
acceleration of the moving mouth, with $\bar{v}$ the corresponding
average velocity, and $\bar{\gamma}=1/\sqrt{1-\bar{v}^2}$ is the
average on the fuzzy relativistic factor; finally $F(\ell)$ is a
form factor that vanishes in the half of the ringhole which is
assumed to be kept motionless, and rises up on average from 0 to 1
as one moves along the direction of the moving mouth. We must
finally point out that any ringhole which is a time machine even
within our universe will also show an defocused two-rings pattern though 
not so blurred perhaps as that corresponding to an inter-universe
ringhole. The phenomenon of defocusing would also take place in
the single ring pattern produced by a wormhole if this behaves
like a time machine or inter-connects two distinct universes.

The image of the system SDSSJ0946+1006, as photographed by Hubble
Space Telescope's Advanced Camera for Surveys [5], shows a
focusing of the two rings that could be compatible with the two
types of ringholes that we have just considered, though it seems
to be resolved enough as for attributing it to the gravitational
effect from a static ringhole staying in our universe. New
findings would become very useful in order to distinguish better
between these two kinds of ringholes.

We finally briefly discuss the odds of finding a macroscopic
ringhole which is kept stable. It was first argued [6] that only
quantum wormholes, and  hence quantum ringholes, with nearly the
Planck size can be stable, with larger tunnelings being violently
destabilized by quantum effects produced by catastrophic particle
creation taking place near the chronology horizons. Actually,
Hawking even advanced his chronology protection conjecture [7] for
wormholes which can also be applied to ringholes, preventing the
appearance of closed timelike curves, so making the universe safe
for historians and free of the occurrence of the kind of phenomena
dealt with in this letter. Thus, neither wormholes nor ringholes
could exist due to these quantum fluctuation instabilities.

However, besides some counter-examples to the Hawking's conjecture
that includes e.g. some compelling argument by Li and Gott [8], it
has been shown [9] that both macroscopic wormholes and macroscopic
ringholes can be stabilized after the coincidence time by the
accelerating expansion of the universe which induces their throat
to quickly growing comovingly to the super-luminal universal
expansion. On the other hand, similarly to as it happens with
wormholes [10], accretion of phantom energy onto the ringholes
should also induce in them a ultra rapid swelling up that would
circumvent the kind of quantum effects considered by Hawking so
that, such as it also happens with their above-mentioned size
increasing which is comoving to the universal expansion, the
destabilizing quantum effects here cannot act in time to destroy
the tunnel during the current speeding-up of the universe.
Therefore, the odds for ringholes to exist and gravitationally act
on the light coming from luminous sources the way we showed before
appear to be good enough in the the context of our accelerating 
universe as for allowing the kind of interpretation considered 
in this letter.

\acknowledgements

\noindent This work was supported by MICINN under research project
no. FIS2008-06332. The author benefited from discussions with C.
Sig\"{u}enza of the {\it Estaci\'{o}n Ecol\'{o}gica de
Biocosmolog\'{\i}a} of Medell\'{\i}n, Spain.

\end{document}